\documentclass[prb,twocolumn,amsmath,amssymb,superscriptaddress, bibnotes]{revtex4}

\usepackage{color}
\usepackage{graphicx}% Include figure files
\usepackage{dcolumn}% Align table columns on decimal point
\usepackage{bm}% bold math

\begin{document}
\newcommand{\red}[1]{\textcolor{red}{#1}}

\title{Nematic transition and highly two-dimensional superconductivity in BaTi$_2$Bi$_2$O revealed by $^{209}$Bi-nuclear magnetic resonance/nuclear quadrupole resonance measurements}

\author{Shunsaku~Kitagawa}
\email{kitagawa.shunsaku.8u@kyoto-u.ac.jp}
\author{Kenji~Ishida}
\affiliation{Department of Physics, Kyoto University, Kyoto 606-8502, Japan}
\author{Wataru~Ishii}
\author{Takeshi~Yajima}
\author{Zenji~Hiroi}
\affiliation{Institute for Solid State Physics, The University of Tokyo, Kashiwa, Chiba 277-8581, Japan}

\date{\today}

\begin{abstract}
In this Rapid Communication, a set of $^{209}$Bi-nuclear magnetic resonance (NMR)/nuclear quadrupole resonance (NQR) measurements has been performed to investigate the physical properties of superconducting (SC) BaTi$_2$Bi$_2$O from a microscopic point of view.
The NMR and NQR spectra at 5~K can be reproduced with a non-zero in-plane anisotropic parameter $\eta$, indicating the breaking of the in-plane four-fold symmetry at the Bi site without any magnetic order, i.e., ``the electronic nematic state''.
In the SC state, the nuclear spin-lattice relaxation rate divided by temperature, $1/T_1T$, does not change even below $T_{\rm c}$, while a clear SC transition was observed with a diamagnetic signal.
This observation can be attributed to the strong two-dimensionality in BaTi$_2$Bi$_2$O.
Comparing the NMR/NQR results among BaTi$_2$$Pn$$_2$O ($Pn$ = As, Sb, and Bi), it was found that the normal and SC  properties of BaTi$_2$Bi$_2$O were considerably different from those of BaTi$_2$Sb$_2$O and BaTi$_2$As$_2$O, which might explain the two-dome structure of $T_{\rm c}$ in this system.
\end{abstract}

\maketitle

%\section{Introduction} %% No sections necessary for express letters, letters and short notes
An electronic nematic transition, which is characterized by spontaneous rotational symmetry breaking in the electronic system is often observed in strongly coupled superconductors\cite{R.Okazaki_Science_2011,S.Kitagawa_PRB_2013,S.-H.Baek_NatMat_2014,S.Hosoi_PNAS_2016,Y.Sato_NatPhys_2017}.
The nematic phase extends beyond the superconducting (SC) phase in the phase diagram of most superconductors.
The relationship between superconductivity and the nematic state, and the origin of the nematic transition, especially in iron-based superconductors, are phenomena of intensive debate in the superconductivity community.
However, the relationship and the origin remain elusive.
Therefore, to understand these phenomena, it is important to investigate and compare various superconductors that exhibit a nematic transition.

\begin{figure}[!tb]
\vspace*{10pt}
\begin{center}
\includegraphics[width=8.5cm,clip]{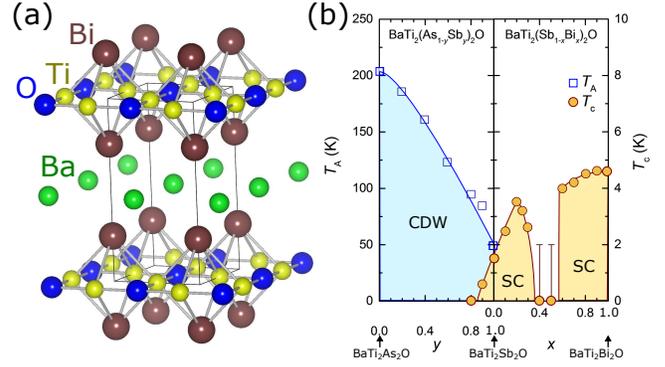}
\end{center}
\caption{(a)Crystal structure of BaTi$_{2}Pn_{2}$O.
The figure of the crystal structure is modeled on the three-dimensional visualization program VESTA\cite{K.Momma_JAC_2011}.
(b) Electronic phase diagram of BaTi$_2$(As$_{1-y}$Sb$_{y}$)$_{2}$O and BaTi$_{2}$(Sb$_{1-x}$Bi$_{x}$)$_{2}$O\cite{T.Yajima_JPSJ_2013}.
SC denotes the superconducting phases. 
The open squares represent the CDW transition temperature $T_{\rm A}$. 
The solid circles represent the superconducting temperature $T_{\rm c}$ determined by magnetization and electrical resistivity measurements.
}
\label{Fig.1}
\end{figure}

Recently, superconductivity in the vicinity of a charge density wave (CDW) phase was discovered in BaTi$_{2}Pn_{2}$O ($Pn$ = As, Sb, and Bi)\cite{T.Yajima_JPSJ_2012,T.Yajima_JPSJ_2013,T.Yajima_JPSJ_2013_2,T.Yajima_CM_2017}, which possesses a two-dimensional layered structure as shown in Fig.~\ref{Fig.1}(a).
BaTi$_{2}Pn_{2}$O crystallizes into a tetragonal structure of space group $P4/mmm$ (No.123, $D_{4h}^{1}$) with alternately stacked Ti$_2Pn_2$O layers and Ba atoms along the $c$ axis.
The Ti$_2Pn_2$O layers contain a Ti$_2$O square net, which is an anti-configuration to the CuO$_2$ square net.
The edge-shared TiO$_{2}Pn_{4}$ octahedra form a square lattice, and the electronic state of Ti$^{3+}$ is in the 3$d^1$ state, which is regarded as an electron--hole symmetric state of the 3$d^9$ state in Cu$^{2+}$\cite{T.Yajima_JPSJ_2012}. 
BaTi$_2$As$_2$O shows an anomaly in the nematic state at $T_{\rm A}$ = 200~K, which is ascribed to a CDW transition\cite{X.F.Wang_JPhys_2010}.
%%%%%%%%%%%%%%%%%%%%%%%%%%% Figure 1 %%%%%%%%%%%%%%%%%%%%%%%%%%%%%%%%%%%%%
%%%%%%%%%%%%%%%%%%%%%%%%%%%%%%%%%%%%%%%%%%%%%%%%%%%%%%%%%%%%%%%%%%%%%%%%%%%
This anomaly was suppressed by the substitution of Sb for As and $T_{\rm A}$ becomes $\sim$40~K in the end member BaTi$_2$Sb$_2$O\cite{S.Kitagawa_PRB_2013}.
In BaTi$_2$Sb$_2$O, an SC transition was also observed at an SC transition temperature of $T_{\rm c}$ = 1.2~K.
$^{121/123}$Sb-nuclear magnetic resonance(NMR)/nuclear quadrupole resonance(NQR) measurements in BaTi$_2$Sb$_2$O revealed the breaking of in-plane fourfold symmetry at the Sb site below $T_{\rm A}$ without an internal field appearing at the Sb site\cite{S.Kitagawa_PRB_2013}, which indicated an electronic nematic transition at $T_{\rm A}$.
The muon spin rotation ($\mu$SR) measurements also indicated no internal magnetic field below $T_{\rm A}$\cite{Y.Nozakia_PRB_2013}, and a long-range structural phase transition was found in the neutron diffraction measurements\cite{A.Benjamin_NatCommun_2014}.
These results can be understood with commensurate nematic CDW ordering\cite{H.Nakaoka_PRB_2016} or $p$-$d$ bond order\cite{D.W.Song_arXiv_2018} below $T_{\rm A}$.
In both cases, the nematic state is realized in the CDW phase.
Moreover, $^{121/123}$Sb-NMR/NQR measurements strongly suggest that SC symmetry is a conventional $s$ wave in BaTi$_2$Sb$_2$O\cite{S.Kitagawa_PRB_2013}, which is in sharp contrast with those in the cuprate and iron-based superconductors.

With the substitution of Bi for Sb, $T_{\rm A}$ was suppressed, $T_{\rm c}$ shows a dome shape with maximum $T_{\rm c}$ = 3.5~K at $x = 0.2$, and superconductivity terminates at $x = 0.4$ in BaTi$_2$(Sb$_{1-x}$Bi$_{x}$)$_2$O\cite{T.Yajima_JPSJ_2013}.
Interestingly, superconductivity reappears upon further substitution, and a two-dome structure in $T_{\rm c}$ was observed as shown in Fig.~\ref{Fig.1}(b).
In the end member BaTi$_2$Bi$_2$O, an SC transition was observed at 4.6~K without any trace of the CDW/spin density wave (SDW) transition.
Since there is a possibility that the SC properties of BaTi$_2$Bi$_2$O are different from those of BaTi$_2$Sb$_2$O, it is important to investigate BaTi$_2$Bi$_2$O to understand the normal and SC properties of the BaTi$_{2}Pn_{2}$O system.

In this Rapid Communication, $^{209}$Bi-NMR/NQR measurements have been performed to investigate the physical properties of BaTi$_2$Bi$_2$O from a microscopic point of view.
From the temperature evolution of the NQR spectra, an electronic nematic transition at $\sim$45~K was discovered.
In the SC state, the nuclear spin-lattice relaxation rate divided by temperature, $1/T_1T$, does not change even below $T_{\rm c}$, while a clear SC transition was observed by an ac susceptibility measurement.
This is ascribed to the strong two-dimensionality in BaTi$_2$Bi$_2$O.
Although the NQR spectra in both BaTi$_{2}$Sb$_{2}$O and BaTi$_{2}$Bi$_{2}$O indicate the breaking of fourfold symmetry at the Sb/Bi site below $\sim$40~K, the normal and SC properties of BaTi$_2$Bi$_2$O are considerably different from those of BaTi$_{2}$Sb$_{2}$O.

%%%%%%%%%%%%%%%%%%%%%%%%%%% Figure 2 %%%%%%%%%%%%%%%%%%%%%%%%%%%%%%%%%%%%%
\begin{figure}[!b]
\vspace*{10pt}
\begin{center}
\includegraphics[width=8.5cm,clip]{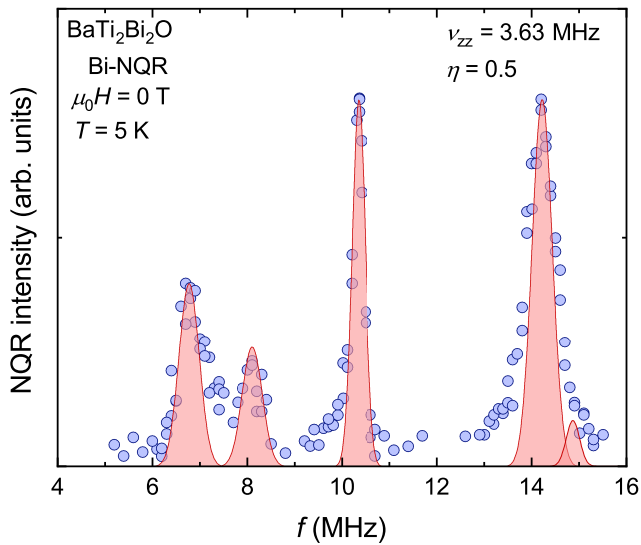}
\end{center}
\caption{$^{209}$Bi-NQR spectra obtained by the frequency-swept method at 5~K.
From the observed $^{209}$Bi-NQR spectra, the quadrupole parameters for Bi nuclei are evaluated as shown in the figure.
The solid red curves are the simulation of the NQR spectra using the estimated quadrupole parameters.
}
\label{Fig.2}
\end{figure}
%%%%%%%%%%%%%%%%%%%%%%%%%%%%%%%%%%%%%%%%%%%%%%%%%%%%%%%%%%%%%%%%%%%%%%%%%%%

%\section{Experimental}
Polycrystalline samples of BaTi$_{2}$Bi$_{2}$O were synthesized via the conventional solid-state reaction\cite{T.Yajima_JPSJ_2013}. 
Stoichiometric amounts of BaO, Ti, and Bi were mixed and pelletized.
The pellet was then wrapped in a Ta foil and sealed in a quartz tube. 
The reaction temperature was 850$^{\rm o}$C.
To prevent sample degradation by air and/or moisture, the polycrystalline samples were mixed with Araldite adhesive.
The mixture was then solidified with random crystal orientation. 
All the procedures were performed in a glove box filled with Ar.
The SC transition at 4.6~K was confirmed by a dc magnetization measurement with a commercial superconducting quantum interference device (SQUID) magnetometer (Quantum Design, MPMS3), and the ac susceptibility measurement was performed using an NMR coil after the mixing.
A spin-echo technique was used for the NMR/NQR measurements.
The $^{209}$Bi-NMR spectra (nuclear spin $I$ = 9/2, nuclear gyromagnetic ratio $^{209}\gamma/2\pi = 6.842$~MHz/T, and natural abundance 100\%
) were obtained as a function of magnetic field in a fixed frequency $f$ = 68.41~MHz ($\sim10$~T).
The $^{209}$Bi nuclear spin-lattice relaxation rate $1/T_1$ was determined by fitting the time variation of the spin-echo intensity after the saturation of the nuclear magnetization to a theoretical function for $I$ = 9/2.

%%%%%%%%%%%%%%%%%%%%%%%%%%% Figure 3 %%%%%%%%%%%%%%%%%%%%%%%%%%%%%%%%%%%%%
\begin{figure}[!tb]
\vspace*{10pt}
\begin{center}
\includegraphics[width=8.5cm,clip]{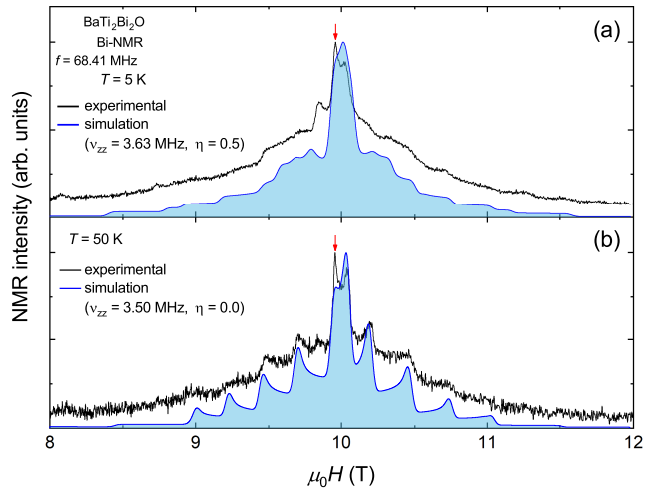}
\end{center}
\caption{Field-swept $^{209}$Bi-NMR spectra at (a) 5~K and (b) 50~K at 68.41~MHz.
In addition, simulations of the NMR spectra are shown.
At 5~K, the same NQR parameters as those for the NQR measurements are used for the simulation.
The arrows indicate the magnetic field at which $T_1$ is measured.
}
\label{Fig.3}
\end{figure}
%%%%%%%%%%%%%%%%%%%%%%%%%%%%%%%%%%%%%%%%%%%%%%%%%%%%%%%%%%%%%%%%%%%%%%%%%%%

%\section{Results and Discussion}
The $^{209}$Bi-NQR measurements indicate an electronic nematic state at low temperatures.
Figure~\ref{Fig.2} shows the $^{209}$Bi-NQR spectra that were obtained by the frequency-swept method at 5~K.
When $I \ge 1$, a nucleus has an electric quadrupole moment $eQ$ as well as a magnetic dipole moment; thus, the degeneracy of nuclear energy levels is lifted even at zero magnetic field due to the interaction between $eQ$ and the electric field gradient (EFG). 
This interaction is described as 
\begin{align}
\mathcal{H}_Q &= \frac{h \nu_{zz}}{6}\left\{(3I_z^2-I^2)+\frac{1}{2}\eta(I_+^2+I_-^2)\right\},
\label{eq.1}
\end{align}
where $h$ is the Planck's constant, $\nu_{zz}$ is the quadrupole frequency along the principal axis ($c$-axis) of the EFG and is defined as $\nu_{zz} \equiv 3e^2qQ/2I(2I-1)$ with $eq = V_{zz}$, and $\eta$ is an asymmetry parameter of the EFG expressed as $(V_{xx}-V_{yy})/ V_{zz}$ with $V_{\alpha \alpha}$, which is the second derivative of the electric potential $V$ and the EFG along the $\alpha$ direction ($\alpha = x,y,z,$).
When $^{209}$Bi is in the EFG, the degenerate ten nuclear-spin states are split into five energy levels, yielding four (or more) resonance frequencies as shown in Fig.~\ref{Fig.2}.
The NQR parameters $\nu_{zz} = 3.63$~MHz and $\eta = 0.5$ were obtained by comparing the observed $^{209}$Bi-NQR spectra and calculated resonance frequencies obtained from the diagonalization of Eq.\eqref{eq.1}.
A nonzero $\eta$ implies the breaking of fourfold symmetry at the Bi site, which has a $C_{4}$ symmetry in the tetragonal structure (also called the ``electronic nematic state'').
It was observed that the low-intensity peak at the highest frequency corresponded to the $\nu_1$ + $\nu_2$, which is caused by the formation of hybrid states due to nonzero $\eta$\cite{K.Karube_JPSJ_2011}.

\begin{figure}[!tb]
\vspace*{10pt}
\begin{center}
\includegraphics[width=7.5cm,clip]{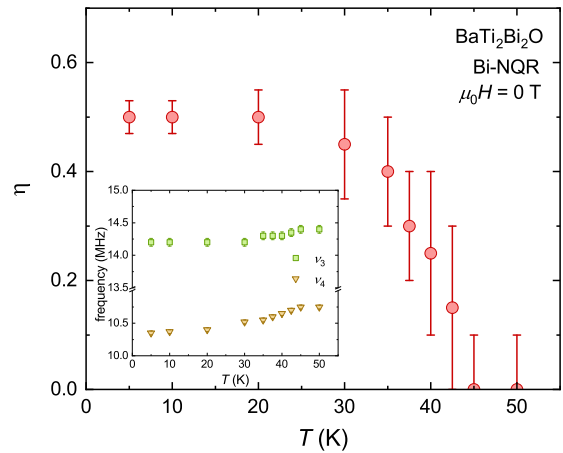}
\end{center}
\caption{Temperature dependence of $\eta$ deduced from the first and second largest NQR peaks ($\sim$10~and 14~MHz).
Inset: Temperature dependence of NQR frequency.
}
\label{Fig.4}
\end{figure}

The nonzero $\eta$ was also confirmed by the field-swept NMR spectrum as shown in Fig.~\ref{Fig.3}.
For the NMR measurements, although the nuclear energy levels were already split by the electric quadrupole interaction, magnetic fields were still applied to lift the degeneracy of the spin degrees of freedom.
The total effective Hamiltonian could then be expressed as
\begin{align}
\mathcal{H} &= \mathcal{H}_{\rm Z} + \mathcal{H}_{\rm Q} \notag \\
            &= -\frac{\gamma}{2\pi} h (1 + K)\bm{I} \cdot \bm{H} + \mathcal{H}_{\rm Q},
\end{align}
where $K$ is the Knight shift and $\bm{H}$ is an external field.
As the NMR spectra are varied against the angle between the principal axis of the EFG and magnetic field direction, the sum of the spectrum for all the corresponding angles is observed in the case of the powder samples.
The NMR spectrum at 5~K was consistently reproduced by the NQR parameters determined through the NQR measurements at 5~K.
In contrast, the NMR spectrum at 50~K can be fitted by the simulation with $\nu_{zz} = 3.50$~MHz and $\eta = 0.0$.
The small difference between experimental and simulated values might have originated from the impurity phase and/or the degree of orientation.
It should be noted that the double-horn-shaped satellite signals are the characteristic feature of the $\eta = 0.0$.
This reflects the preservation of the fourfold symmetry of the crystal structure.
To estimate the transition temperature, the temperature evolution of the NQR spectra was measured.
Figure~\ref{Fig.4} shows the temperature dependence of $\eta$ deduced from the first and second largest NQR peaks ($\sim$10~and 14~MHz).
The figure indicates that $\eta$ assumes a nonzero value below $T_{\rm anom}~\sim$45~K.
As such, a nematic phase transition was not reported in previous experiments\cite{T.Yajima_JPSJ_2013_2}.
The most plausible reason behind this is that the anomalous transition may have been so small that it could be detected only by a highly sensitive probe in an electric environment, such as the NQR measurement.
A similar nematic transition was observed at $\sim$40~K in BaTi$_2$Sb$_2$O.
To reveal the origin of the nematic transitions and the mechanism by which the nematic phases cover the two SC domes in BaTi$_2$(Sb$_{1-x}$Bi$_{x}$)$_2$O, it is important to first understand the relationship between superconductivity and the electronic nematic state.

%%%%%%%%%%%%%%%%%%%%%%%%%%% Figure 4 %%%%%%%%%%%%%%%%%%%%%%%%%%%%%%%%%%%%%
%%%%%%%%%%%%%%%%%%%%%%%%%%%%%%%%%%%%%%%%%%%%%%%%%%%%%%%%%%%%%%%%%%%%%%%%%%%

%%%%%%%%%%%%%%%%%%%%%%%%%%% Figure 5 %%%%%%%%%%%%%%%%%%%%%%%%%%%%%%%%%%%%%
\begin{figure}[!tb]
\vspace*{10pt}
\begin{center}
\includegraphics[width=8.5cm,clip]{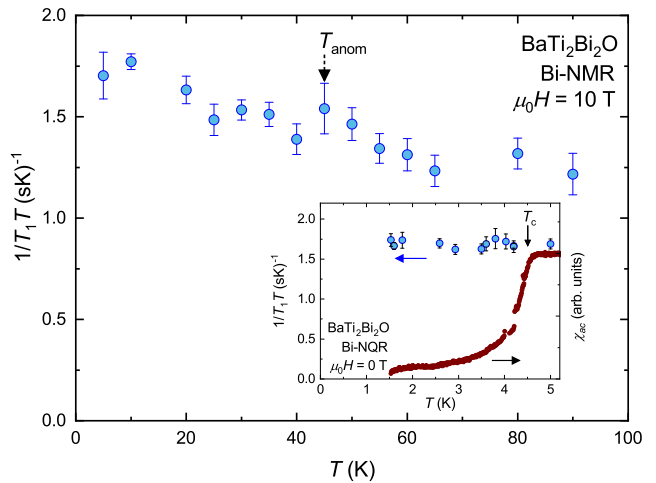}
\end{center}
\caption{Temperature dependence of $1/T_1T$ at $\sim$10~T.
The dashed arrow indicates the temperature at which the anomaly was observed in the NMR spectrum $T_{\rm anom}$.
Inset: Temperature dependence of $1/T_1T$ and ac susceptibility at 0~T.
The solid arrow indicates the SC transition temperature $T_{\rm c}$.
}
\label{Fig.5}
\end{figure}
%%%%%%%%%%%%%%%%%%%%%%%%%%%%%%%%%%%%%%%%%%%%%%%%%%%%%%%%%%%%%%%%%%%%%%%%%%%

Figure~\ref{Fig.5} shows the temperature dependence of $1/T_1T$ at $\sim$10~T.
$1/T_1T$ was measured at the center peak of the Bi-NMR spectrum as indicated by arrows in Fig.~\ref{Fig.3}.
$1/T_1T$ slightly increases upon cooling and no measurable anomaly was observed around $T_{\rm anom}$.
To investigate the SC properties, the temperature dependence of $1/T_1T$ at the NQR signal (14.2~MHz peak) as shown in the inset of Fig.~\ref{Fig.5} was also measured.
While a clear SC transition was observed at 4.6~K by the ac susceptibility measurement, $1/T_1T$ did not change even below $T_{\rm c}$.

First, we discuss the unaltered behavior of $1/T_1T$ in the SC state.
In conventional superconductors, $1/T_1T$ decreases at temperatures well below $T_{\rm c}$ due to the reduction of the quasi-particle density of states around the Fermi energy $E_{\rm F}$ by opening the SC energy gap.
There are two possible reasons why $1/T_1T$ did not decrease in the SC state.
One possibility is that spin diffusion due to low dimensionality prevented the reduction of $1/T_1T$\cite{M.-H.Julien_JPSJ_2008}.
A similar absence of a clear reduction of $1/T_1T$ was observed in (La$_{0.87}$Ca$_{0.13}$)FePO\cite{Y.Nakai_PRL_2008} and certain cuprate superconductors\cite{S.Kambe_PRL_1994,M.-H.Julien_PhysicaB_2003}.
This brings us to the other possibility that the coupling between the Ti$_{2}$O layer and Bi atom was so weak that the effect of the SC transition could not be detected even by $^{209}$Bi-NQR measurements.
Such a scenario was discovered at the Cu(1) site in YBa$_{2}$Cu$_{3}$O$_{7-\delta}$ and YBa$_{2}$Cu$_{4}$O$_{8+\delta}$\cite{M.Mali_PLA_1987,H.Zimmermann_PhysicaC_1989}.
$1/T_1$'s at the first and second largest NQR peaks (10.35 and 14.2~MHz) are not different, although the frequency dependency is expected in the spin diffusion scenario\cite{M.-H.Julien_JPSJ_2008}.
In both cases, low dimensionality is important for such an anomalous behavior to occur; therefore, it can be concluded that the NMR results indicate that BaTi$_{2}$Bi$_{2}$O is a two-dimensional superconductor, and $T_{\rm c}$ in this system is enhanced by low dimensionality\cite{Y.Yanase_PhysRep_2003}.
Such two dimensionality was not predicted by earlier band calculations\cite{D.V.Suetin_JETPL_2013,K.Nakano_SciRep_2016}, which imply three-dimensional Fermi surfaces.
It is plausible that this discrepancy is related to the electronic nematic transition.
The electronic state at low temperatures differs from that at room temperature due to the nematic transition at $\sim$45~K.

%%%%%%%%%%%%%%%%%%%%%%%%%%% Figure 6 %%%%%%%%%%%%%%%%%%%%%%%%%%%%%%%%%%%%%
\begin{figure}[!tb]
\vspace*{10pt}
\begin{center}
\includegraphics[width=8.5cm,clip]{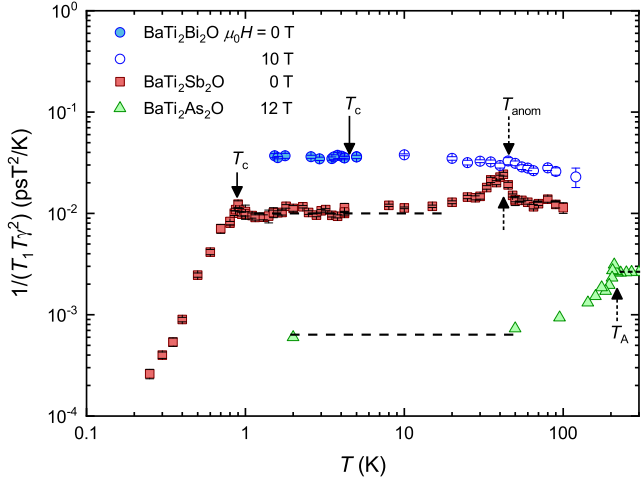}
\end{center}
\caption{(Color online)Temperature dependence of $1/T_1T$$1/(T_{1}T\gamma^2)$ in BaTi$_{2}$Bi$_{2}$O, BaTi$_{2}$Sb$_{2}$O\cite{S.Kitagawa_PRB_2013}, and BaTi$_{2}$As$_{2}$O\cite{D.W.Song_arXiv_2018}.
For BaTi$_{2}$Bi$_{2}$O, we plot the data at 10~T above 5~K and at 0~T below 5~K.
For BaTi$_{2}$As$_{2}$O, we use the angle average of $1/T_1T$ for simplicity.
The dashed arrows indicate $T_{\rm anom}$ or $T_{\rm A}$.
The solid arrows indicate $T_{\rm c}$.
}
\label{Fig.6}
\end{figure}
%%%%%%%%%%%%%%%%%%%%%%%%%%%%%%%%%%%%%%%%%%%%%%%%%%%%%%%%%%%%%%%%%%%%%%%%%%%

Next, we compare the NMR/NQR results in BaTi$_{2}$Bi$_{2}$O with those in BaTi$_{2}$Sb$_{2}$O and BaTi$_{2}$As$_{2}$O.
In general, $1/T_1T$ can be described as
\begin{align}
\frac{1}{T_1T} = \frac{2\gamma_N^2k_{\rm B}}{(\gamma_e^2\hslash)^2}\lim_{\omega_0 \rightarrow 0}\sum_{\bm{q}}A_{\bm{q}}A_{-\bm{q}}\frac{\chi"_{\perp}(\bm{q},\omega_0)}{\omega_0},
\end{align}
where $\gamma_N$ ($\gamma_e$) is the nuclear (electronic) gyromagnetic ratio, $k_{\rm B}$ is the Boltzmann's constant, $\hslash$ is the reduced Planck's constant, $A_{\bm{q}}$ is the Fourier transform of the hyperfine coupling, and $\chi''_{\perp}(\bm{q},\omega)$ is the transverse component of the imaginary part of the dynamical susceptibility.
Then, the value of $1/T_1T$ in different nuclei may be compared after the normalization by $\gamma_N^2$.
Figure~\ref{Fig.6} shows the temperature dependence of $1/(T_{1}T\gamma_N^2)$ in BaTi$_{2}$Bi$_{2}$O, BaTi$_{2}$Sb$_{2}$O, and BaTi$_{2}$As$_{2}$O\cite{D.W.Song_arXiv_2018}.
While the NQR/NMR spectra in those compounds indicate the breaking of the fourfold symmetry at the $Pn$ site, the temperature dependences of $1/T_1T$ are considerably different from each other.
In BaTi$_{2}$Sb$_{2}$O and BaTi$_{2}$As$_{2}$O, $1/T_1T$ was clearly enhanced toward $T_{\rm A}$.
The ratio of $1/T_1$ between the two isotopes of the Sb nuclei is close to the ratio of the square of the nuclear gyromagnetic ratio $\gamma_{\rm N}$, suggesting that this enhancement originates from the magnetic nature\cite{S.Kitagawa_PRB_2013}.
In addition, a reduction of the constant value of $1/T_1T$ below $T_{\rm A}$ was observed, indicating a decrease in the density of states due to the nematic transition.
The value of $1/(T_{1}T\gamma_N^2)$ in BaTi$_{2}$As$_{2}$O is one order of magnitude smaller than those in BaTi$_{2}$Sb$_{2}$O and BaTi$_{2}$Bi$_{2}$O, possibly due to the small hyperfine coupling constant.
In contrast, $1/T_1T$ shows no measurable anomaly around $T_{\rm anom}$ in BaTi$_{2}$Bi$_{2}$O.
While the system does exhibit strong two dimensionality, the value of $1/T_1T$ in BaTi$_{2}$Bi$_{2}$O is larger than those in BaTi$_{2}$Sb$_{2}$O and BaTi$_{2}$As$_{2}$O.
This observation can be attributed to the existence of strong coupling between the atoms of Bi and Ba and the lack thereof between the Bi and Ti$_{2}$O layers.
From this, it can hence be inferred that the BaBi$_2$ layer in BaTi$_{2}$Bi$_{2}$O plays the role of a block layer whose electronic state is different from that of the Ti$_{2}$O plane.
%%%%

Furthermore, a clear coherence peak immediately below $T_{\rm c}$ and an exponential decay at low temperatures as evidence of an $s$-wave superconductivity were observed in BaTi$_{2}$Sb$_{2}$O, while no anomaly was observed in BaTi$_{2}$Bi$_{2}$O.
These differences may have arisen from the difference of the dimensionality between two compounds.
Since BaTi$_{2}$Sb$_{2}$O is three dimensional, the anomaly in the nematic transition and superconductivity can be detected by $^{121/123}$Sb-NMR/NQR.
However, such an anomaly was not observed by $^{209}$Bi-NMR/NQR in BaTi$_{2}$Bi$_{2}$O because of its strong two dimensionality.
It is apparent that the SC Ti$_{2}$O layer is sandwiched by the nonsuperconducting BaBi$_{2}$ block layer with the different electronic state; thus, BaTi$_{2}$Bi$_{2}$O is regarded as a two-dimensional superconductor.
Therefore, it is important that the validation of two-dimensional superconductivity is performed using $^{47/49}$Ti or $^{17}$O NMR measurements.

%\section{Conclusion}
In conclusion, the $^{209}$Bi-NMR/NQR measurements were performed to investigate the physical properties of superconducting BaTi$_2$Bi$_2$O from a microscopic point of view.
The temperature evolution of the NQR spectra indicates an electronic nematic order at $T_{\rm anom} \sim$45~K.
Comparing the NMR/NQR results among BaTi$_2$$Pn$$_2$O ($Pn$ = As, Sb, and Bi), it was found that the normal and SC properties of BaTi$_2$Bi$_2$O were somewhat different from those of BaTi$_2$Sb$_2$O and BaTi$_2$As$_2$O.
The observed two-dome structure in $T_{\rm c}$ on BaTi$_2$(Sb$_{1-x}$Bi$_{x}$)$_2$O may have originated from these differences.

\section*{Acknowledgments}
The authors acknowledge S. Yonezawa, Y. Maeno, and Y. Matsuda for fruitful discussions. 
This work was partially supported by the Kyoto University LTM Center and Grant-in-Aids for Scientific Research (KAKENHI) (Grants No. JP15H05882, No. JP15H05884, No. JP15K21732, No. JP15H05745, No. JP15K17698, and No. JP17K14339). 

%\bibliographystyle{apsrev4-1}
%\bibliography{Ref,NMR}

\end{document}